\documentclass[11pt,twoside]{article}

\usepackage{asp2004}
\usepackage{epsf}
\usepackage{psfig}
\usepackage{lscape}
\usepackage{natbib}
\bibpunct{(} {)} {;} {a} { , } {,} 
\begin{document}
\title{The internal structure of the asteroids}   
\author{V.\v Celebonovi\'c}   
\affil{Inst.of Physics,Pregrevica 118,11080 Zemun-Beograd,Serbia }    

\begin{abstract} 
Knowledge of the internal structure of the asteroids is rather poorly developed,mainly due to lack of reliable data (masses or densities and radii). This contribution has two aims: to review the basics of existing knowledge and explore the possibility of advancing the field by introducing and using in it some well known ideas from solid state physics.
\end{abstract}


\section{Introduction}

The largest part of astronomical research directly relies on data taken at various kinds of telescopes of variable size - from those having mirrors of 8 or 10 meters in diameter to backyard amateur telescopes. However,serious comlications arise when one attempts to gain information on the internal structure of any kind of celestial bodies.

The reason for these complications is simple: the interiors of {\bf any} kind of objects are unobservable. For example, nobody has ever gone (and nobody ever will) into the center of Jupiter to measure the temperature there. Surfaces of many kinds of objects are observable. However, the "state" of a surface of any given object is a consequence of at least two factors: processes occuring in the interior of the object but also of any possible external influences. 

The aim of this lecture is to discuss to some extent one particular "subfield" within the field of modelling of the internal structure of celestial objects - the internal structure of the asteroids. Apart from the introduction,this contribution has three more parts. In the following section a review will be given of the present knowledge on the internal structure of the asteroids. In particular,a {\it vocabulary} of the definitions of the most important notions  used in the literature on this topic will be presented. Knowledge on the interiors of the asteroids will be discussed there from the point of view of astronomy. The third section is devoted to a discussion  of possible applications of  solid state physics to studies of the interiors of the asteroids. The fourth and final section contains the conclusions.             

\section{Asteroid interiors-the astronomical aspect of the problem}
\subsection{Preliminaries}
The first asteroid,1 Ceres,has been discovered on the first night of the $XIX$ century - January $1$,$1801$ by Father Giuseppe Piazzi from the Astronomical Observatory in Palermo. The number of discoveries slowly rose throughout the $XIX$ century while only visual techniques were used in searches. Around $1890$ Max Wolf in Wienna introduced photography in asteroid observation,which accelerated the pace of new discoveries. The total number of objects listed in the {\it Ast-Dys} data base kept at the Department of Mathematics of the University of Pisa (Italy) is at the time of this writing (beginning of April 2006) around {\bf a quarter of a milion} objects. To be precise,on April $10$ $2006$ there were 129436 numbered and 112511 unnumbered asteroids. 

Studies of asteroids are at present performed for two strong  motives. Fundamentally,
it assumed in the research community that the asteroids represent nearly unchanged primordial material from the earliest phases of the existence of our planetary system. Studying them gives the opportunity to get knowledge about the origin and chemical composition of the protoplanetary disk,its changes with heliocentric distance,various mixing processes...  

There exists also a "practical" motivation. Namely, orbit calculations have shown that there  
exists a small group of objects called "potentially hazardous asteroids" (PHA). There are at present $777$ known PHA objects. In their motion they can come closer than $0.05$ A.U. to the Earth or even less (which amounts to nearly $20$ Earth-Moon distances). It is not impossible  than one of these objects collides with the Earth, provoking various dammage and destruction. The amount of destruction resulting from such a colision would be a function of the kinematical parameters of the motion of the object,its {\bf chemical composition} and the geographical coordinates of the impact point. 
\subsection{The problem}
We have now encountered the first purely observational problem of modern astronomy related to asteroid interiors. High resolution observations show that most asteroids have irregular shapes,as if they were outcomes of a catastrophic collision. The importance of collisions in asteroid studies is demonstrated by the existence of families-groupings of asteroids with nearly identical orbits. For a recent review of asteroid family indentifications see,for example,(Bendjoya and Zappala,2002). How does one determine the mean mass density of an asteroid without assuming the sphericity of its shape?  Masses can be determined from asteroid,planetary and spacecraft perturbations. The volume can be determined from direct immaging from ground based observatories and/or spacecraft,and the bulk density can then be calculated in a simple way. This procedure has been applied to the biggest and best observed asteroids,and some of the results are presented in the following table (Hilton 2002). 


\begin{table}[!ht]
\caption{Mean densities of some asteroids}
\smallskip
\begin{center}
{\small
\begin{tabular}{ccccc}
\tableline
\noalign{\smallskip}
Name & Density & \\
          &[kg m$^{-3}$]& \\
\noalign{\smallskip}
\tableline
\noalign{\smallskip}
1 Ceres& $2060\pm50$\\
2 Pallas& $3100\pm30$\\
4 Vesta& $3500\pm20$\\
45 Eugenia& $1200\pm600$\\
433 Eros &$2670\pm30$\\
\noalign{\smallskip}
\tableline
\end{tabular}
}
\end{center}
\end{table}
 
Taking into account the error bars,it follows that the mass density of asteroid $45$ Eugenia may be {\bf lower} than the density of water. Another interesting example of a low density asteroid is $617$ Patroclus. Its density has recently been determined as $\rho=800^{+200}_{-100}$ kg m$^{-3}$ (Marchis et.al,2006),which is again lower than the density of water. 
These two asteroids are examples of a class of asteroids called "rubble piles" - they consist of rocks loosely bound together,with gaps between them. It is usually assumed that "rubble pile" asteroids are a result of catastrophic collisions and subsequent reaccumulation of the debris. For a recent review of the subject of catastrophic disruption of the asteroids see for example (Michel, Benz and Richardson,2004). 

A line of study aiming at drawing some conclusions on the interior of the asteroids is the analysis of their shapes. This demands extremely high resolution data,avaliable at just a handful of facilities throughout the world. An interesting example concerns $1$ Ceres. Using  data avaliable in the middle of the eighties,and a particular theoretical framework,the present author has concluded that this asteroid is undifferentiated (\v Celebonovi\'c,1987). However,a recent analysis of HST data shows that there are strong indications that $1$ Ceres is differentiated 
(Thomas,Parker,McFadden et al.,2005). Another asteroid with differentiated interior is $4$ Vesta (for example Keil,2002). 

Studying the internal structure of small bodies opens the possibility of getting answers to several important scientific questions (Binzel,A'Hearn,Asphaug et al.,2003). Briefly speaking,some of these can be formulated as follows:

\begin{enumerate}
	\item The limiting size at which planetary processes determine the internal structure. Planetary processes "operate" in the big planets,and allow for some degree of differentiation of their interiors. On the other hand,in the small bodies,such as the asteroids,dominate material characteristics and deffects,which are completely irrelevant on planetary scale. The transition between the two regimes is estimated to occur in objects of the diameter of $\approx1000$ km (Binzel,A'Hearn,Asphaug et al.,2003). Roughly speaking,this is the diameter of $1$ Ceres .One should note here that the possibility of differentiation is also composition dependent;a small asteroid made up of material with higher atomic mass will have more chances of having a differentiated interior. 
	\item Which objects are "survivors" from the formation stages of the planetary system? Such objects would provide a sort of "windows" to the shapes and internal structures created in accretion. It is almost certain the two such asteroids exist; these are Ceres and Vesta. However,identification of the smallest asteroids which survived since accretion is an open problem.
	\item What is the most probable size of rubble piles and monoliths? A detailed answer to this question depends on the interplay between material strength and gravitation.It is virtually certain that all the smaller asteroids have been shattered by collisions. This problem involves a chain of issues related to material science so it is completely open to study form the viewpoint of solid state physics.

\end{enumerate}

In this section we have briefly reviewed the  possibilities of "pure astronomy"  in getting knowledge on the interiors of the asteroids. The remaining part of this contribution is devoted to the possibilites of solid state physis, both theoretical and experimental,in getting information on this subject.  

\section{Asteroid interiors-the physical aspect of the problem}

For a physicist,an asteroid is a solid body of known orbit,mass,volume,density,
albedo,reflection spectrum. Starting from the reflection spectrum it is possible,in principle,to derive the chemical composition of the surface. Numerically,this means that (in principle) one can obtain the mean atomic mass $A$ of the surface layer,the ionic charge $Z$ and the ionic mass $M$ of the surface of the object.

Observations show that collisions played an important role in the evolution of the planetary system,so at the beginning we have to introduce definitions of several notions.         

\subsection{The vocabulary} 

\begin{itemize}
	\item {\bf Stress} is the ratio of the force applied on a body to the cross section of the surface of a body normal to the direction of the force.
\end{itemize}
\begin{itemize}
	\item {\bf Strain} is the geometrical expression of the action of stress. 
\end{itemize}
\begin{itemize}
	\item {\bf Material strength} is the capacity of a material to withstand stress and strain.
\end{itemize}
\begin{itemize}
	\item {\bf Porosity} is the ratio of the volume of all the pores on the object to the volume of the object.
\end{itemize}
\begin{itemize}
	\item {\bf Tensile Strength} is the force required to pull a material object to the point where it breaks up.
\end{itemize}
\begin{itemize}
	\item {\bf Fragmentation} is the separation of a body into pieces under the action of stress. 
\end{itemize}
These definitions are in a sense "borrowed" from solid state physics. Pertaining to asteroids proper,one frequently encounters the following notions:
\begin{itemize}
	\item {\bf Monoliths} are asteroids consisting of one single block of material formed in the early stages of the Solar System.
\end{itemize}
\begin{itemize}
	\item {\bf Agregates} are asteroids made up of material of increased porosity compared to the monoliths.
\end{itemize}
\begin{itemize}
	\item {\bf Shattered} asteroids are those whose interiors 
are dominated by joints and cracks. 
\end{itemize}
\begin{itemize}
	\item {\bf Rubble piles} are asteroids which were shattered (by colisions) and afterwards reassembled. 
\end{itemize}

\subsection{The calculations and experiments}
Standard solid state physics (such as Landau and Lifchitz,1976) gives the following expression for the lattice energy per unit volume of a crystal
\begin{equation}
(E-N\epsilon_{0})/V	= \frac{\pi^{2}(k_{B} T)^{4}}{10 (\hbar \bar{u})^{3}}
\end{equation}
Equating this energy with the kinetic energy per unit volume exchnaged in a collision,one could get an estimate of the kinetic energy needed for complete destruction of an asteroid. All the symbols in eq.(1) have their standard meanings,and $\bar{u}$ is the mean velocity of sound in the material. This quantity is related to the equation of state (EOS) of the solid making up the asteroid by
\begin{equation}
\bar{u}^{2}= \frac{\partial P}{\partial \rho}
\end{equation}
where $P$ and $\rho$ denote the pressure and the mass density of the material. As the aim is to relate asteroid studies as much as possible to solid state physics,the obvious question is how can one measure the pressure dependence of the value of $\bar{u}$ . 

The instrument used in most high presure experiments in the last 50 years is the diamond anvil cell (DAC),presented on the following figure taken from (Jayaraman,1983). 
\begin{figure}[!h]
\plotone{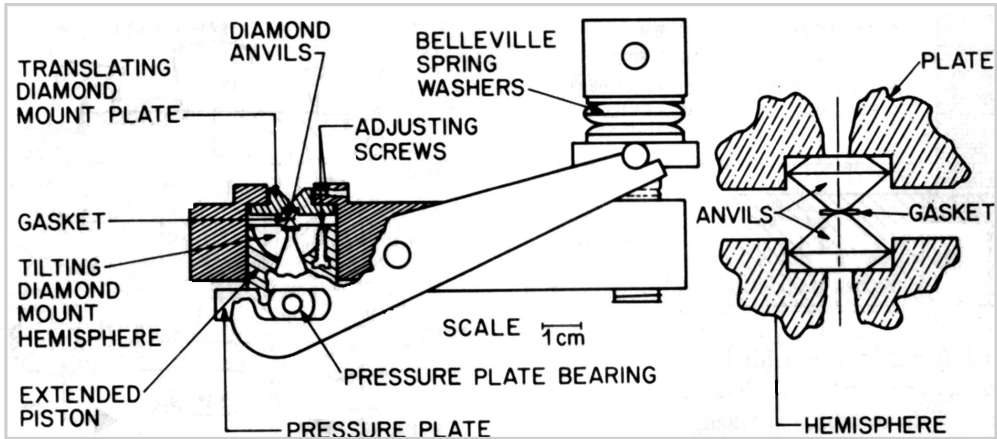}
\caption{Cross section of a NBS type DAC}
\end{figure}

The basic principles of functioning of a DAC are discussed in the literature.A recent example is (\v Celebonovi\'c,2006). Basically,a specimen a hydrostatic and a pressure sensor are inserted in a small hole drilled in a thin metal plate (called {\it the gasket}) which is inserted between the diamonds. The specimen is excited by an incoming laser beam,and its deexcitation spectrum is recorded at different values of external pressure. Due to the small sizes of the diamonds,the hole in the gasket is small,and so are the specimen and the pressure sensor. Filling a DAC (that is,inserting the specimen,pressure sensor and some hydrostatic,in the hole in the gasket) is extremely delicate. A DAC can be used to experimentally determine the equation of state of any given material (in asteroid studies the limitation is obviously to solids). The equation of state is usually expressed in the form of an expression linking the pressure $P$ and the density $\rho$, so knowing it would give the possibility to determine the mean velocity of sound,and by eq.(1) the lattice energy per unit volume.Equating this to the kinetic energy exchanged in a collision would give the kinetic energy needed for a complete destruction of an asteroid of given chemical composition. In principle,such a calculation can be performed,but its result would be experimentally useless.

A much more interesting calculation would be a determination of the energy needed to produce a fracture in a material of given chemical composition relevant for the asteroids.The basic aim of such a calculation would be the fracture stress - this is "simply" the stress needed to make a fracture in a given material. A small crack which for some reason occurs in a material will increase the stress in the region of the crack tip compared to the stress in the region far from the crack. A fracture in a material  will occur when the stress at the tip of the crack becomes equal to the "cohesive stress" (Tilley,2004). It can be shown that a fracture in a material occurs when the stress is equal to
\begin{equation}
	\sigma_{c}=\frac{1}{2}(\frac{E\gamma\rho}{r_{0}a})^{1/2}
\end{equation}
In this expression $E$ denotes Young's modulus of the material,$\gamma$ is the surface energy,$\rho$ is the radius of curvature of the crack tip,$r_{0}$ is the interatomic spacing at which stress is equal to zero and the length of the crack is $2a$.

This equation is seemingly simple. However,each of the quantities which enter into it is in itself a result of a separate calculation of a different degree of complexity.
The calculation of the Young modulus starts from the expression for the lattice potential energy,which has the following form:
\begin{equation}
U_{L}= \frac{- C_{1}}{r}+\frac{C_{2}}{r^{n}}	
\end{equation}
where $C_{1}$ and $C_{2}$ are constants. The interatomic force is defined as $F=-\frac{\partial U}{\partial r}$ and it can be shown (Tilley,2004) that the Young modulus is given by
\begin{equation}
E=\frac{1}{r_{0}}(\frac{\partial F}{\partial r})_{r=r_{0}}=\frac{C_{1}(1-n)}{r_{0}^{4}}	
\end{equation}
The symbol $r_{0}$ denotes the separation between the atoms in the equilibrium configuration. 

Calculating the  surface energy of any given material is a complicated task. Studying the occurence of fractures in different types of asteroids,and connecting them to the collisions which produced them is,astronomically speaking,a very interesting subject. However,in advance of it, a series of calculations aiming at determining all the quantities in eq.(3) would have to be performed. This series of calculations is completely in the domain of solid state physics,and (in a sense) opens the door to an interplay of solid state physics and "pure" astronomy in asteroid studies. 
\subsection{Differentiation in the interior}
There exist different proofs that planets in our Solar System have a certain number of layers in their interiors. For a recent review of the subject see,for example, Stacey (2005). The origin of these layers is basically ascribed to the influence of high pressure and/or temperature on the behaviour of materials in deep planetary interiors. Related to the asteroids,the obvious question is are there physically reliable chances that differentiation occurs in their interiors? 

Quantum mechanics,laboratory and geophysical experiments  show that high external pressure changes the electronic structure of atoms and molecules making up any material. 

Observation (in principle) gives the mass,density and surface chemical composition of an asteroid. A simple calculation can give an estimate of the central pressure in the object.  
Using the elementary definition of hydrostatic pressure,the central pressure in an asteroid of mass $M$ and radius $R$ can be estimated as
\begin{equation}
	P=\rho\gamma\frac{M}{R}
\end{equation}
On the other hand, quantum mechanics combined with statistical physics gives the possibility of estimating the pressure needed for changing the electronic structure of any given atom or molecule. In this way,one can theoretically estimate whether or not an asteroid made up of a material of mean atomic mass $A$ can be expected to be differentiated.
The distribution of the pressure,density and temperature with depth can be obtained by any of the existing equations of state of a solid under high pressure. 

High temperature can also change the electronic structure of atoms and molecules. Physical reasoning shows that in collisions of asteroids transformation of a part of the kinetic energy of their relative motion into heat was certainly occuring. Both asteroids participating in a collision would certainly heat up,the exact amount depending on their heat capacity. As a sort of "experimental example" it suffices to recall the recent impact of the Deep Impact probe into the nucleus of comet Tempel 1.In this event the space probe disintegrated and actually evaporated,while the comet nucleus was heated and shaken. 
\subsection{Impacts} 

In this lecture we have so far discussed various aspects of  gaining fundamental knowledge about the asteroids and,by analyzing them, on formation of the planetary system. There is also a very practical motive for the study of the asteroids,and this is the possibility that some of them impact into the Earth and cause  dammage of variable extent. Of course,the possibility of collision with the Earth exists also for comets or parts of asteroids. Dammage occuring in such impacts could be of variable extent,ranging from localized consequences (for example in the case of a meteorite falling on a house) to a global catastrophe. Such impacts have occured in history,and the best known and most recent example is the "Tunguska" event from 1908 when a small asteroid exploded at a height of approximately $6-10$ kilometers ower the Tunguska region of Siberia. 

There exist now  nearly 800 asteroids for which the probability of impacting into the Earth is not identically zero. Current data on this class of objects are avaliable at {\it http://neo.jpl.nasa.gov}.  Studying their motion,predicting their close encounters with the Earth and estimating the probability of impact in such passages is a problem in the domain of pure astronomy. Hazards concerning the possibility of a NEO asteroid impacting the Earth have recently been discussed (Champan,2004). In the remaining of this contribution we shall try to give some hints on what solid state physics and material science can contribute to the study of asteroid impacts into the Earth. 

For a material scientist,the collision of an asteroid with the Earth can be simply described as the impact of a solid impactor into a bilayered target;the first layer of the target is a fluid (the atmosphere)  and the second one is either fluid or solid (that is the ocean or the continent). 

An asteroid approaches the atmosphere with certain velocity. The upper layers of the atmosphere,which it hits at first,are dense compared to the virtually empty interplanetary space. Therefore, the asteroid heats up. 
The consequences of this heating depend on the heat capacity of the material of which the asteroid is composed,the speed and angle with which it hits the atmosphere and the density of the atmosphere. 

The heat capacity of a solid at low temperature is given by (Landau and Lifchitz,1976)
\begin{equation}
	C=\frac{2\pi^{2}(k_{B}T)^{3} V}{5 (\hbar \bar{u})^3}
\end{equation}
 where  $\bar{u}$ is the mean velocity of sound in the interior of the asteroid,related to the equation of state of the material making up the asteroid and $V$ is the volume of the asteroid. 

When it enters the atmosphere,the asteroid "brakes". The resulting change of kinetic energy is spent on heating and forming cracks and fractures in the asteroid. Equating the change of the kinetic energy of the asteroid to the sum of its heat capacity and the energy needed to form cracks and fractures in it, gives the possibility to determine what will happen to the asteroid. That is,will it hit the surface unchanged,breakup into several pieces which will then hit the surface,or will it burn up in the atmosphere.

If the incoming asteroid burns in the atmosphere,the only consequence on the surface will be the occurence of a shock wave,which will cause some localized dammage. The exact amount of dammage will depend on the speed,mass and chemical composition of the incoming object, and on the nature of the surface. A real "experimental example" of such an event occured in 1908 in the Tunguska region of Siberia. The shock wave which appeared destroyed the forest at a surface of more than $2000$ km$^{2}$. No impact crater exists on the spot where the explosion occured. A small number of craters  resulting from impact of asteroids in various geological periods has been identified on the Earth ( Kharif and Pellinovsky,2005). A more recent example of the impact of  a comet into a planet was widely observed when comet Shoemaker-Levy $9$ impacted into Jupier and left traces in its atmosphere. One can only immagine the consequences of such an impact into the Earth.  

The situation is much more complicated in the case when the asteroid (either intact or broken into pieces) hits the surface of the Earth. Determining the characteristics of possible impactors is a complex long term task for astronomy but also solid state physics. Astronomers can "provide" the optical spectrum of an object. However,transiting from the observed spectrum to the chemical composition and (for example) material strength is a problem for chemistry and  solid state physics. There have been proposals to establish a data bank of physical and chemical data on possible impactors ( Huebner and Greenberg,2001). 

If the impactor (or possibly a "chain of impactors") hits the solid surface of the Earth,it will certainly provoke heavy dammage on the spot  where the impact occurs. An interesting question (again for solid state physicists) is to determine the characteristics of a possible crater which will occur as a consequence of the impact. These calculations are extremely complex,because they have to take into account a multitude of factors: the velocity,impact angle, chemical composition,internal structure and porosity of the impactor,but also the chemical composition,porosity and other mechanical parameters of the soil. For a recent example of a theoretical study of crater formation  taking into account the influence of porosity see (Wunnemann,Collins and Melosh 2006).

There have been attempts of experimental studies of impacts of small mass projectiles on large impacts. Such experiments were often performed several decades ago,but for completely different motives (for example,Maurer and Rinehart,1960). 

Finally,if an asteroid imacts into any of the oceans,a tsunami will occur,the details depending on the multitude of parameters of the asteroid but also on the geographical coordinates of the impact.
\section{Conclusions}    
In this contribution we have discussed the problem of gaining knowledge on the internal structure of the asteroids.  The main difficulty in this field is scarcity  data on the masses,volumes and spectra of the asteroids. Some known astronomical results were reviewed,and in particular a brief vocabulary of terms used in this field was presented. Attempts were made throughout the text to indicate possibilities for people engaged in solid state and/or material science research to enter asteroid studies,because there exist numerous possibilities for their engagement.

\section{Acknowledgement}
The author is grateful to Professor Osman Demircan and the local organizing comitee for organizing a very interesting meeting in the beautiful setting of the Ankara University Summer Campus . I am grateful to Dr.Zoran Kne\v zevi\'c from the Astronomical Observatory in Belgrade for helpful comments about a previous version of this manuscript. The preparation of this contribution was financed as a part of the research project 141007 of the Ministry of Science and Protection of the Environement of Serbia.


\end{document}